\documentclass[]{aa} 

\usepackage{graphicx}

\usepackage{txfonts}
\begin{document}

   \title{Three-dimensional magnetic reconnection in a collapsing coronal loop system}


   \author{Aidan M. O'Flannagain\inst{1} 
                \and Shane A. Maloney\inst{1}
        \and Peter T. Gallagher\inst{1}
        \and Philippa Browning\inst{2}
        \and Jose Refojo\inst{3}}

   \institute{Astrophysics Research Group, School of Physics, Trinity College Dublin, Dublin 2, Ireland
         \and
            Jodrell Bank Centre for Astrophysics, University of Manchester, Manchester, M13 9PL, UK
          \and
                        Trinity Centre for High Performance Computing, Trinity College Dublin, Dublin 2, Ireland
                         }

   \date{Received October 18, 2017; accepted June 18, 2018}

\authorrunning{O'Flannagain, Maloney, Browning, Gallagher, Refojo}
\titlerunning{Reconnection in a collapsing coronal null}

 
  \abstract
   {Magnetic reconnection is believed to be the primary mechanism by which non-potential energy stored in coronal magnetic fields is rapidly released during solar eruptive events. Unfortunately, owing to the small spatial scales on which reconnection is thought to occur, it is not directly observable in the solar corona. However, larger scale processes, such as associated inflow and outflow, and signatures of accelerated particles have been put forward as evidence of reconnection.}
   {Using a combination of observations we explore the origin of a persistent Type I radio source that accompanies a coronal X--shaped structure during its passage across the disk. Of particular interest is the time range around a partial collapse of the structure that is associated with inflow, outflow, and signatures of particle acceleration.}
   {Imaging radio observations from the Nan\c{c}ay Radioheliograph were used to localise the radio source. Solar Dynamics Observatory (SDO) AIA extreme ultraviolet (EUV) observations from the same time period were analysed, looking for evidence of inflows and outflows. Further \texttt{mpole} magnetic reconstructions using SDO HMI observations allowed the magnetic connectivity associated with the radio source to be determined.}
   {The Type I radio source was well aligned with a magnetic separator identified in the extrapolations. During the partial collapse, gradual (1 km/s) and fast (5 km/s) inflow phases and  fast (30 km/s) and rapid (80--100 km/s) outflow phases were observed, resulting in an estimated reconnection rate of $\sim$0.06. The radio source brightening and dimming was found to be co--temporal with increased soft x-ray emission in both Reuven Ramaty High Energy Solar Spectroscopic Imager (RHESSI) and  Geostationary Operational Environmental Satellite (GOES).}
   {We interpret the brightening and dimming of the radio emission as evidence for accelerated electrons in the reconnection region responding to a gradual fall and rapid rise in electric drift velocity, in response to the inflowing and outflowing field lines. These results present a comprehensive example of 3D null-point reconnection.}


   \maketitle
%

\section{Introduction}
Solar eruptive events (SEEs) are believed to be triggered by the release of non-potential magnetic energy stored in twisted coronal magnetic field structures. The primary mechanism proposed for the transfer of this stored energy into the accelerated particles, heated plasma, and ejected matter that make up the SEE is thought to be magnetic reconnection. Proposed in the 1940s, and built upon by \citet{swe58}, \citet{par57}, and \citet{pet64}, it has been demonstrated that, with a suitably small-scale diffusion region, reconnection can release enough energy to drive the SEE within a standard flare time frame \citep{pri00}. Theory has been further developed to include break-up of the reconnecting current sheet into smaller magnetic islands, potentially allowing even greater rates \citep{kli00,shi01}. Reconnection has been reproduced in laboratories \citep{yam10, zho10} and been measured in the magnetosphere of the Earth \citep{oie01, den01}.

\begin{figure*}
\centering
        \includegraphics[width = \textwidth]{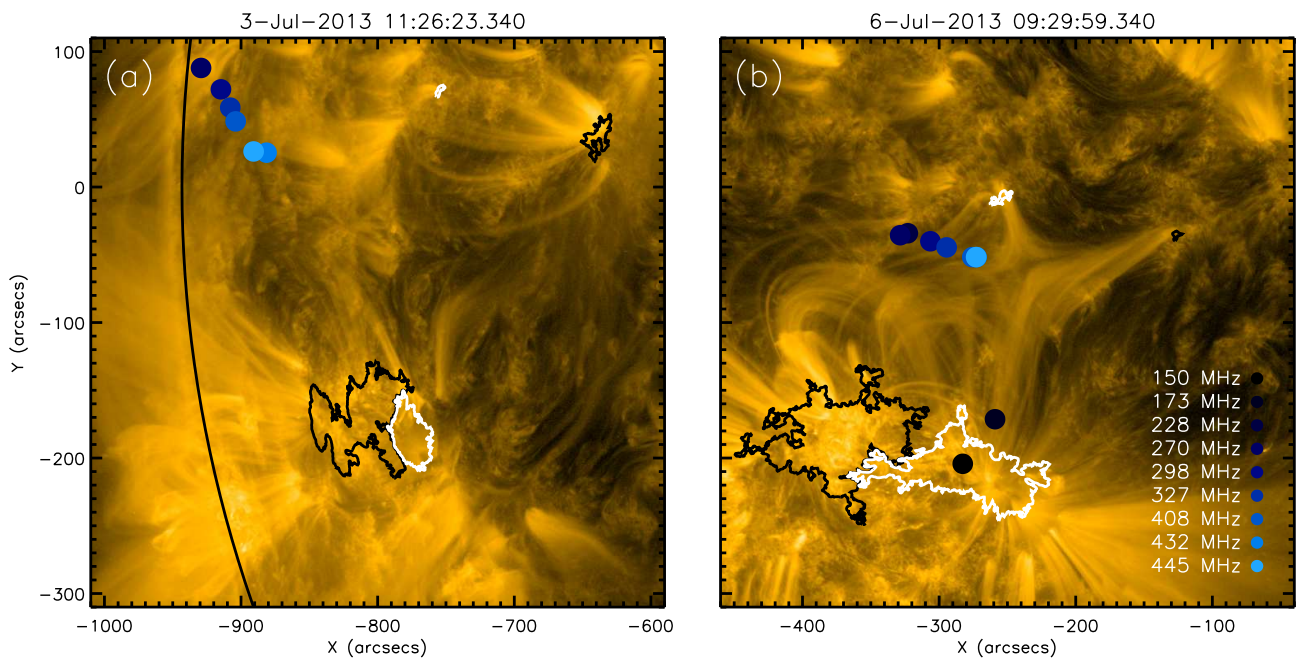}
        \caption{Evolution of active region AR11785 during its approach towards disk centre. AIA 171~\AA~images are shown at times when the active region is (a) near to the eastern limb and (b) near to disk centre. The location and polarity of the underlying line-of-sight magnetic field from HMI are shown in black for negative polarity, and white for positive polarity. These contours represent 80~G in the associated HMI images. Overlaid as blue circles are the location of Type I noise storm source centroids at all available frequencies from 150 to 445~MHz.}
        \label{fig:diskpassage}
\end{figure*}

However, as fast reconnection requires a diffusion region with a spatial scale of the order of the electron gyro-radius \citep{dau12}, it is not directly observable in the solar corona by modern instrumentation. Indirect observations are possible however as reconnection is thought to have certain signatures visible at larger spatial scales. For example, active region loops that appear to be moving in towards and outwards from apparent X-points on the solar limb have been interpreted as reconnection inflow and outflow \citep{sav12, su13}. Inflows have also been observed at the limb along with microwave emission, which were presented as evidence for particle acceleration at an X-point \citep{nar14}. Fast reconnection should also produce super-magnetosonic outflows resulting in termination shocks; these shocks appear in radio data as solar type II radio-like features \citep{Chen}. Separately, Type I radio noise storms have been explored as a signature of accelerated particles at reconnection sites near to the interface between active regions \citep{wil02, wil05}. While these storms are known to occur over very long timescales \citep[e.g.][]{del11}, they have also been observed to react rapidly to flare and coronal mass ejection (CME) activity \citep{aur90, kat07, iwa12}.

In order to address these disparate forms of both non-flaring and impulsive acceleration, we must take into account models that address the 3D nature of reconnection. Three-dimensional reconnection has been divided into a number of different categories, such as spine and fan reconnection \citep{pri96, dal05}, null-point reconnection \citep[e.g.][]{mas09}, and reconnection along quasi-separatrix layers \citep{dem96, dem97}. Simulations of these regions have been shown to produce the accelerated electron distributions expected from flares \citep{pri09, bro10, sta12, bau13}. In particular, it has been shown that reconnection and acceleration are likely along 3D separators, the lines of intersection between separatrices, or surfaces of different magnetic connectivity \citep{gor89, lon05, par10}

In this paper, we present observations of a coronal X--shaped structure that appears near to disk centre on 6 July 2013. This structure persists for the full passage of the active region across the disk, and is closely accompanied by a persistent radio Type I noise storm. In particular, observations of a partial collapse of this structure are discussed, and the implications for coronal magnetic reconnection are explored.

\section{Observations}
The focus of this work is the evolution of a quadrupolar X-shaped coronal structure located to the north or north-west National Oceanic and Atmospheric Administration (NOAA) active region AR11785, identifiable in the coronal channels of Solar Dynamics Observatory (SDO) AIA (94, 131, 171, 193, 211, and 335~\AA) for its full passage across the solar disk, on 2 -- 12 July 2013. For the majority of this time, a Type I radio noise storm is also observable by the Nan\c{c}ay Radioheliograph \citep[NRH;][]{ker97}, consistently located above the X-shaped structure.

An overview of the approach of AR11785 is given in Figure \ref{fig:diskpassage}. Two AIA 171~\AA~images are shown, at two different times during the passage of the active region towards disk centre, covering a total of three days. The X-shaped structure appears to begin forming in panel (a), but is most clear at disk centre (b). The centroids of the NRH radio noise storm sources -- defined as all emission above 50\% of the maximum for each image -- are denoted by blue circles; lighter shades of blue correspond to higher frequencies of radio emission. As shown, higher frequency emission is consistently located closer to the apparent 3D null point.

As Type I noise storms are believed to be plasma emission \citep[e.g.][]{wil08,iwa12}, this would indicate that the higher frequency sources are originating from regions of higher density, and are therefore likely to be lower in altitude above the solar surface. This scenario fits the interpretation of Type I radio noise storms as a column of radially stratified dense plasma above an active region containing some population of accelerated particles \citep{mcl81}. The linear nature of the centroid locations indicates a line of such plasma, which appears to move from north to south over the passage of the active region, whilst still remaining rooted at the 3D null point.

\begin{figure*}
\centering
        \includegraphics[width = \textwidth]{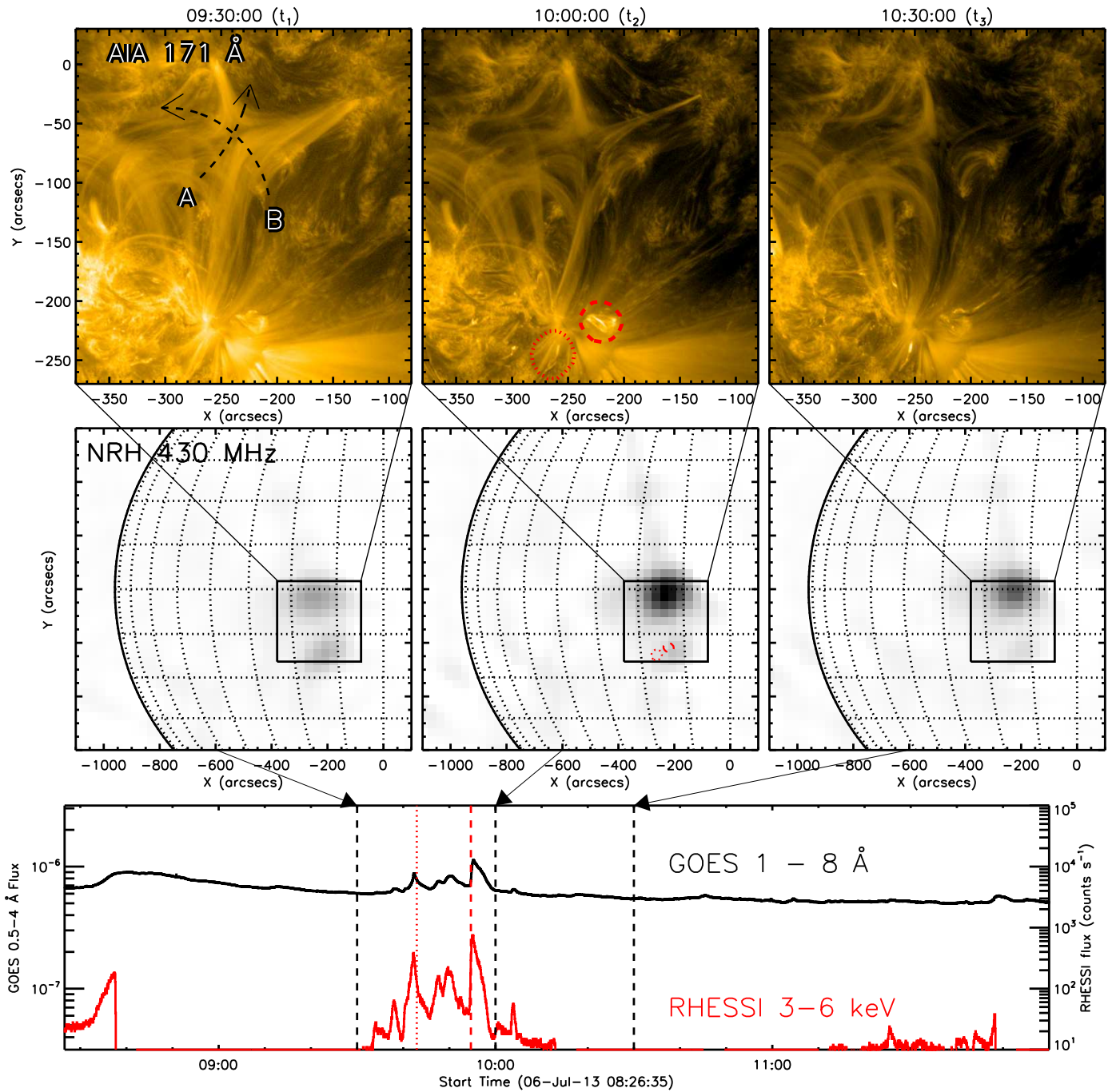}
        \caption{Overview of the collapse of the X-shaped coronal structure. \textit{Top:} AIA 171~\AA~images covering the 90 minutes surrounding the collapse. Overlaid on the earliest image are the slices used to generate the time--distance plots shown in Figure \ref{fig:collapse}. Slices A and B are 67~arcseconds and 96~arcseconds long, respectively; the arrow denotes the reference direction for the time--distance plots. \textit{Middle:} NRH 430~MHz images showing the Type I storm source at the times of the above images. Overlaid is the field of view of the AIA images. \textit{Bottom:} GOES 1--12~keV and RHESSI 3--6~keV light curves, showing the flares that occurred during the collapse. An animation of the collapse in AIA 171~\AA~ data is available on-line.}
        \label{fig:collapse}
\end{figure*}

On 6 July, as the active region and accompanying coronal loop structure approached disk centre, it appeared to undergo a partial collapse beginning at $\sim$09:29~UT. This collapse was concurrent with an enhancement of NRH emission and a series of small C-class flares, which were detected by both the Ramaty High Energy Solar Spectroscopic Imager \citep[RHESSI:][]{lin02} and the Geostationary Operational Environmental Satellite (GOES), which measures spatially integrated solar soft X-rays (SXRs) in 1--12, and 3--25~keV energy bands. A summary of this collapse is shown in Figure \ref{fig:collapse}. Prior to the collapse, four loop structures, connecting each opposite-polarity footpoint pair, are visible in the AIA 171~\AA~channel at $\sim$09:29 UT. One hour later, after the collapse and flare have occurred, the northern and western segments of the X-shaped structure are no longer visible in the 171~\AA~image, and partially diminished in 193~\AA.

The brightening NRH source (seen in 430~MHz images in Figure \ref{fig:collapse}, third row) roughly maintained its overall position to the top left of the coronal 3D null point. Meanwhile, RHESSI imaging demonstrates that hard X-rays (HXRs) originated from the apparent bottom right leg of the quadrupolar structure visible in AIA 171~\AA. Indicating that energy release occurred during this time, perhaps as a result of the destabilisation of the coronal structure. However, as the purpose of this work is to identify the reconnection event itself, hereafter we focus on the coronal location of the collapse and radio noise storm, first by estimating the magnetic field.

\section{Data analysis}
\subsection{Potential field extrapolation}
\begin{figure*}[t!]
\centering
        \includegraphics[width = \textwidth]{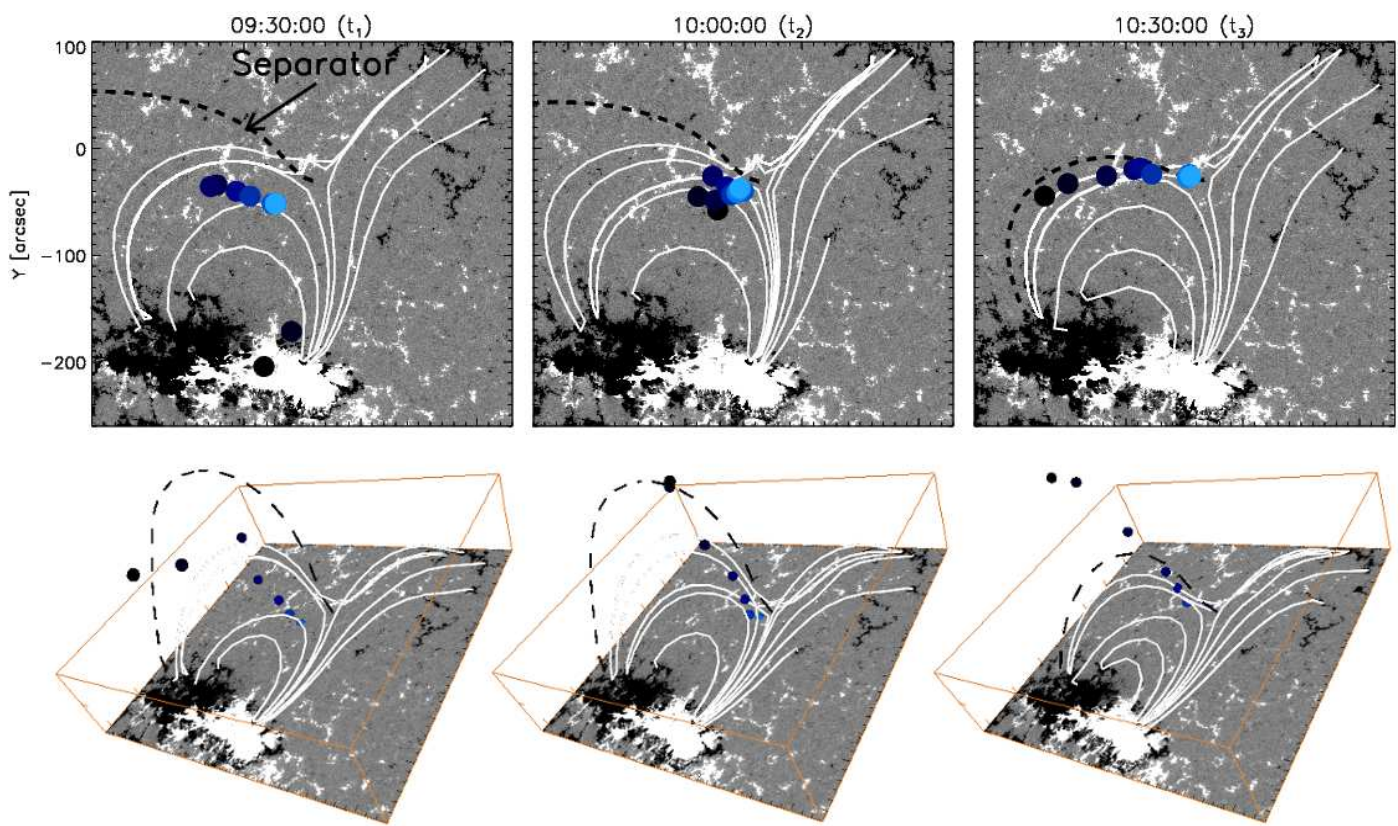}
        \caption{Results from the \texttt{mpole} potential field extrapolations. \textit{Top:} Samples of extrapolated potential magnetic field lines (white), viewed from above, overlaid on the HMI line-of-sight magnetogram produced before (\textit{left}), during (\textit{middle}), and after (\textit{right}) the collapse of the 3D null point. Blue filled circles lie on the locations of NRH centroids, and the colours match the correlation to frequency given in Figure \ref{fig:collapse}. Overplotted onto each frame as a black dashed line is the location of the separator that passes closest to the NRH centroids. The three times were chosen to match those shown in Figure \ref{fig:collapse}. \textit{Bottom:} The same magnetic field lines and radio sources shown in the top row are shown again, viewed at an angle to represent the 3D structure of the field. The dashed line again refers to the separator. Radio source heights are estimated directly from the one--fold Newkirk model \citep{new61}. An animated version of this figure is 
available on-line.}
        \label{fig:mag}
\end{figure*}

In order to interpret these observations in the context of magnetic reconnection, the structure of the coronal magnetic field at the location of the active region must first be estimated. To accomplish this potential field line extrapolations were obtained using the the \texttt{mpole} suite of Interactive Data Language (IDL) programs \citep{lon01,lon02, lon04}. This software simplifies the line-of-sight magnetic field, usually produced by a magnetogram such as those provided by HMI, into a number of positive and negative poles. Treating each region of radial field as magnetic point charges enables us to define the 3D surfaces separating magnetic field lines of differing connectivity (sepatratrices) and the 3D lines of intersection between these surfaces (separators) \citep{lon96}. The presence of the strong current sheets expected at these separators identify these sheets as important locations of 3D reconnection \citep{par10}, occurring, for example, between an emerging and existing active region \citep{lon05}. Therefore it could be expected that signatures of accelerated particles, such as the NRH emission observed in this paper, could be located along these separators.

The magnetic field topology produced by these extrapolations is shown in Figure \ref{fig:mag}. As shown, the extrapolation indeed produces an X-shaped system of magnetic fields, where a 3D coronal null point was found, although the detailed structure differs from that observed in AIA 171~\AA. The null point was located via analysis of the extrapolation results in conjunction with the AIA and radio observations. This topology appears to remain roughly constant throughout the collapse phase, despite the clear outflow and dimming of the observed western loops shown in Figure \ref{fig:collapse}. This is because the input magnetograms do not change significantly over this time period. During the extrapolation process, a large number of separators were produced ($\sim$400--500 per time interval) and so for comparison only the separator that most closely passes the NRH sources (overplotted in blue) is shown for each time interval. However, the majority of separators did originate at the 3D null point.

The relationship between the separator and the line of NRH sources is of particular interest in these results. As shown, prior to the collapse, a separator does appear to pass through the area occupied by the NRH sources, although almost perpendicularly to their apparently linear formation. At the second time interval, shortly after the last X-ray flare occurs, the sources are gathered closer to the 3D null point, and are marginally better aligned with the separator. However, by the third time interval 30~minutes later the sources (with increasing frequency) are pointed directly towards the 3D null point, well-aligned with the separator originating from it.

The heights of the radio sources are estimated directly from a one-fold Newkirk model \citep{new61}. It is often necessary to scale such density models based on density sensitive diagnostics to obtain accurate heights for specific events. In this case no such supporting observations were present, additionally the density of the radio emitting plasma is expected to change throughout the event. As such the heights are not quantitatively accurate but instead give a qualitative representation.

The mpole method is based upon the assumption of discreet isolated photospheric sources; this is an approximation to the real photospheric magnetic field distribution on the Sun. Observations of even the quietest regions of the photosphere show weak vertical fields. The assumption of isolated sources means that many more separators are produced than would be using a more realistic continuous flux distribution. However in the continuous case analogues to the mpole separators have been found in quasiseparatrix layers \citep{dem96} or hyperbolic flux ropes \citep{tit02}. In the limit that the photospheric field becomes discrete it can be shown these features become the separators in the  \texttt{mpole} extrapolations. As such caution must be used when choosing separators in \texttt{mpole} extrapolations as not all separators have  counter parts under the more realistic continuous case.

As these extrapolations are produced using the assumption of a potential force-free field, it is expected that they do not produce an accurate model of magnetic field line geometry for solar active regions shortly before flares, as this is when a large amount of non-potential energy is stored \citep[e.g.][]{wie05, mur13}. It follows that if this non-potential energy is released to drive the observed flares, the coronal field structure should become more potential after energy release adding weight to the idea that the NRH sources lie along the 3D separator. As non-potential magnetic energy is lost from the active region, these NRH sources come to lie along the separator produced from the potential extrapolation, as this is, following the flare, a more accurate model of the coronal field.

\subsection{Extreme ultraviolet and radio comparison}

\begin{figure*}
\centering
        \includegraphics[width = \textwidth]{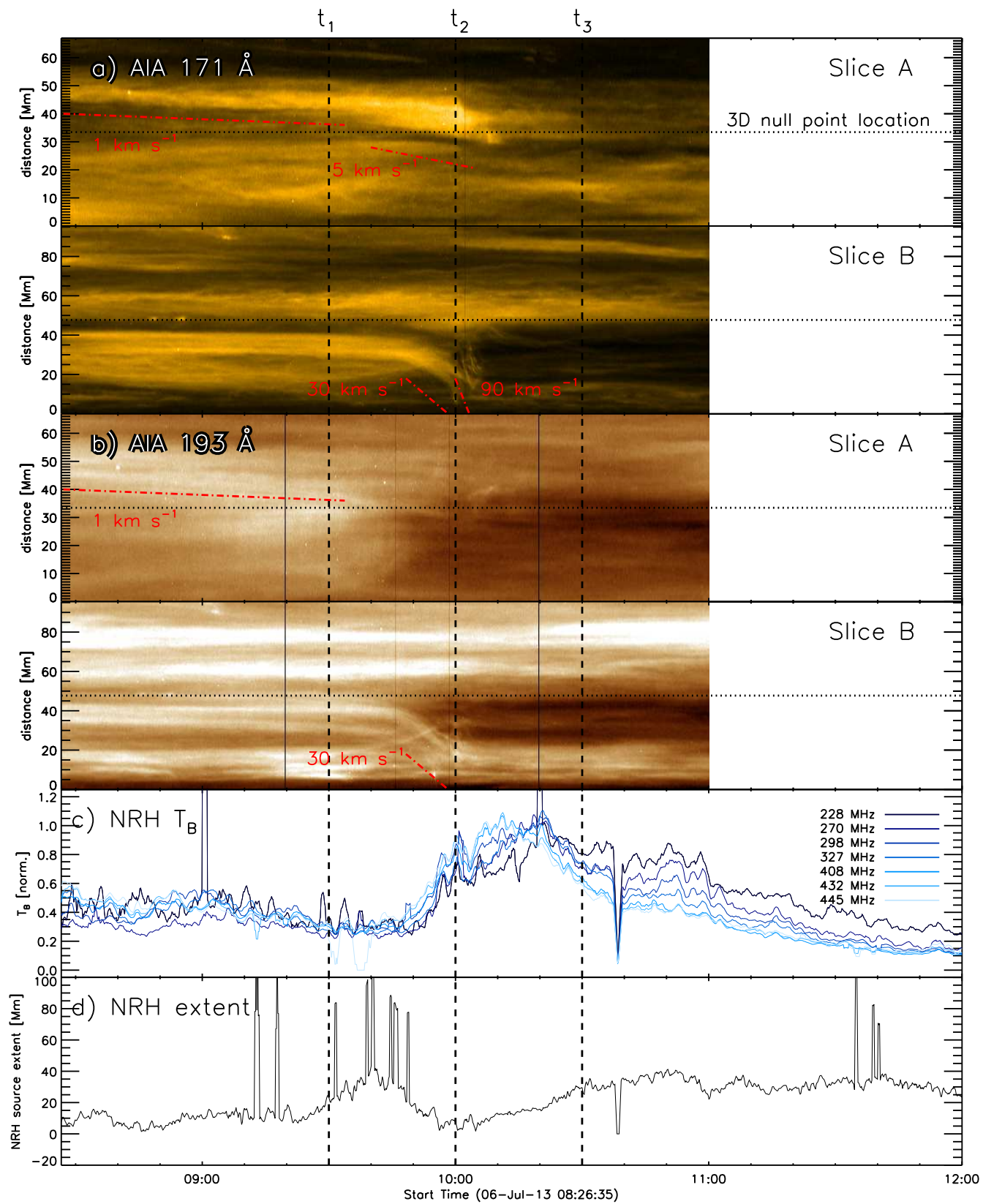}
        \caption{\textit{(a)} Time--distance plots demonstrating the apparent plane-of-sky motion of brightening in the 171~\AA~channel along slice A (top panel) and slice B (bottom panel), which are shown in Figure \ref{fig:collapse}. The horizontal dotted line denotes the location of the apparent coronal 3D null point, and vertical dashed lines correspond to the three time intervals used to produce the three panels in Figure \ref{fig:collapse}. \textit{(b)} Time--distance plots for the 193~\AA~channel. Overlaid on both time-distance plots are dash-dotted lines indicating approximate velocities of the flow features from which they are offset. \textit{(c)} NRH brightness temperature, averaged over the area within a 50\% contour of each NRH image, for its 7 highest recorded frequencies, smoothed over 6 points, and normalised to clarify local peaks. \textit{(d)} NRH extent, measured as the linear distance between the centroid of the highest and lowest frequency stable sources, which were 432 and 298~MHz, respectively.}
        \label{fig:flowplots}
\end{figure*}

In order to characterise the inflow and outflow at the 3D null point, AIA 171 and 193~\AA~flux values were recorded along two curved slices, as shown in Figure \ref{fig:collapse}. The flux along these slices is shown against time for the AIA 171 and 193~\AA~channels in the first two panels of Figure \ref{fig:flowplots}. As shown, there is evidence in both channels of a gradual inflow along slice A from $\sim$08:30 to 10:00~UT, followed in the 171~\AA~channel by a faster apparent inflow. Notably, the inflow seems to occur earlier in the 193~\AA~channel. From 09:40 to 10:10~UT, a much more rapid outflow is apparent along slice B and a faster outflow period of finer loop strands appears shortly after 10:00~UT. In both channels, the gradual inflow exhibited a velocity along the slice of $\sim$ 1~km/s with the faster inflow in 171~\AA~of $\sim$5~km/s. The fast outflow shortly before 10:00~UT had a velocity of $\sim$30~km/s in both channels and  the rapid outflow reached $\sim$80--100~km/s.

These velocities can be used to measure the magnetic reconnection rate in this event. Taking the faster inflow as $V_{in}$ and the rapid outflow of finer loop strands as our $V_{out}$, the reconnection rate $M_{A} = V_{in}/V_{A} \approx V_{in}/V_{out}$ can be estimated, where $V_{A}$ is the Alfv\'en speed. Using this approximation, the reconnection rate was found to be $M_{A} =$ 0.06, agreeing well with the lower rate value reached in the event studied in \citet{su13}.

Of particular interest is the timing of the brightening and motion of the NRH radio sources during the collapse. Shown in panel (c) of Figure \ref{fig:flowplots} are the smoothed and normalised brightness temperatures for all of the NRH observable frequencies, excluding the two lowest. Because these low-frequency sources were not localised to one point for the full duration of the collapse. The given brightness temperatures were acquired by averaging the pixel values over the source area, defined as the 50\% contour of the maximum value for each image. The light curves are smoothed over 6 points for clarity, and normalised to their own maxima, which reached values of $\sim$10, 0.9, 0.4, 0.2, 0.07, 0.04, and 0.03 $\times 10^{9}$~K for frequencies 228, 270, 298, 327, 408, 432, and 445~MHz, respectively. The extent of the NRH sources is shown in the final panel (d) of Figure \ref{fig:flowplots}. This property was produced in an attempt to estimate the length of the NRH source across all frequencies, and so was simply produced by taking the distance between the centroid of the 432 and 298~MHz emission. Local sharp peaks in the plot are produced when images corresponding to one or both of these frequencies are briefly dominated by noise or instrumentational artefacts. 

As shown, during the gradual inflow phase of the collapse, all of the presented NRH channels exhibit a gradual decrease in relative brightness temperature, reaching a minimum around 9:40~UT. Concurrent with a peak in the NRH extent, which corresponded to a spreading of the NRH source frequencies further apart, and further to the solar north-east. This is followed by the simultaneous occurrence of the rapid AIA inflow/outflow, a sudden brightening in all NRH frequencies, and a decrease in the radio source extent. In particular, all NRH channels appear to peak in brightness sharply at 10:00~UT, at the same time as the minimum in source extent. Which is followed by an overall peak in the NRH high-frequency channels, and a subsequent peak in the lower channels roughly 10~minutes later. The NRH emission then decays gradually to background levels, still localised to the same location near to the 3D null point.

\section{Interpretation}
The observations are interpreted in terms of modulation of the reconnection rate at the 3D null point. There are numerous theoretical models describing how the reconnection rate can be modulated by altering the guide fields, reconnecting fields, or electric fields. These specific observations can be better understood by examining the numerical results of the reconnection model outlined in \citet{bro10}. In this work, 2D reconnection theories are built upon by examining particle acceleration at a 3D null point. The primary purpose is to characterise the acceleration for different particle species and for varying electric and magnetic field strengths. An important result of this model is that for stronger magnetic fields, efficiency of particle acceleration was diminished. It is put forward that this is an effect of the electric drift speed having an inverse dependence on magnetic field -- for stronger magnetic fields, a lower drift rate prevented ambient electrons from entering the acceleration region.

While it is also important to consider the effect of variation in the electric fields in the active region, our observational results provide strong evidence supporting the above interpretation. During the gradual inflow phase prior to 09:40~UT, the magnetic field is gathered around the null point with no corresponding observations of an outflow. According to the above model, this increase in field strength would reduce the rate of null-point acceleration, resulting in fewer accelerated particles available to produce non-thermal Type I emission. Conversely, the rapid outflow after 09:40~UT rapidly reduces the magnetic field strength around the null point, allowing for an equivalently rapid increase in the number of accelerated particles and therefore a rapid increase in radio emission across all observed frequencies. The transition from a low reconnection rate with no outflows to a faster reconnection rate with rapid outflows may be driven by strong unobserved electric fields. Once the outflows are present the reduction of magnetic field strength should lead further increases in the reconnection rate.

The behaviour of the NRH source length also supports this picture. The lengthening of the series of NRH sources before 09:40~UT could result from an overall increase of the density along the separator. For a power-law or exponential density structure, a uniform increase in density would shift sources at each frequency upwards and further apart from one another. The rapid decrease in ambient density associated with the reconnection outflow after 09:40~UT would then have the opposite effect, rapidly decreasing the extent of the NRH emission. It should of course be noted that these measurements were taken from plane-of-sky NRH centroids and thus could also have a contribution from 3D motion of the acceleration region -- perhaps the separator itself.

\section{Conclusions}
A solar radio noise storm was identified that exhibited a close consistent association with a quadrupolar transequatorial loop structure for its full passage across the disk. The interpretation of this noise storm as plasma emission produced indirectly by accelerated electrons at a 3D magnetic null point supports the idea that boundaries between interacting or trans-equatorial active regions are ideal locations for non-flaring reconnection due to their opposite magnetic polarity orientations \citep{she75, tsu96, pev00}. This idea of non-flaring reconnection has previously been put forward as a solution to the coronal heating problem \citep{pri03, par07}.

Beyond this long-term relationship, we have investigated the behaviour of the NRH sources during a rapid collapse of the 3D null-point structure, as observed in AIA. We interpreted the variation in radio brightness temperature as a fall and subsequent rise in accelerated electron population caused by the change in electric drift rate as magnetic fields are swept into and then out of the acceleration region. Our interpretation builds a complete picture of both non-flaring and impulsive reconnection by combining observations of coronal inflows and outflows at 3D null points with an associated signature of accelerated particles. As such, this work strongly supports the 3D reconnection models outlined in \citet{dal05} and \citet{bro10}.

This work could be built upon by taking into account the non-potential nature of the interacting active region during the time coming up to the collapse and flare. In this work, a form of potential field extrapolation was used, while it has been shown on many occasions that linear or non-linear force-free extrapolations are better suited to active region magnetic fields \citep{sch08,wie05}. With these types of extrapolations, it could be determined with confidence whether the origin of the Type I radio noise storm was indeed consistently originating at a 3D null.

\begin{acknowledgements}
  We thank Prof. Lidia van Driel-Gesztelyi for her valuable input on this paper. This work has been supported by a Government of Ireland Studentship from the Irish Research Council (IRC) and by ESA/PRODEX.
\end{acknowledgements}


\end{document}